\pdfoutput=1 
\documentclass[cits]{JINST}

\title{A hadronic calorimeter with Glass RPC as sensitive medium.}

\author{G. Grenier$^a$\thanks{On behalf of CALICE collaboration.}\\
\llap{$^a$}IPNL/CNRS-IN2P3/Universit\'e Lyon1\\
  4, rue Enrico Fermi, 69622 Villeurbanne Cedex, France\\
E-mail: \email{grenier@ipnl.in2p3.fr}}

\abstract{The SDHCAL technological prototype is a 
$1 \times 1 \times 1.3$~m$^3$ high-granularity Semi-Digital Hadronic 
CALorimeter using Glass Resistive Plate Chambers as sensitive medium. 
It  is one of the two HCAL 
options considered by the ILD Collaboration to be proposed
for the detector of the future International Linear Collider project.
The prototype is made of up to 50 GRPC detectors of 1~m$^2$ size and 
3~mm thickness each with an embedded semi-digital electronics readout
that is autotriggering and power-pulsed. The GRPC readout is finely segmented into pads of 1~cm$^2$. Measured performances
of the GRPC and the full SDHCAL prototype in terms of homogeneity,
low noise and energy resolution are presented in this proceeding.
}


\begin{document}

\section{Introduction}\label{sec:intro}
This article describes a hadronic calorimeter prototype. 
The Semi-Digital Hadronic CALorimeter (SDHCAL) is 
one of the two hadronic calorimeter options considered by the 
ILD (International Large Detector) Collaboration~\cite{Behnke:2013lya} 
to be proposed for the detector of the future 
International Linear Collider (ILC). ILC detectors are designed for 
Particle Flow Algorithms~(PFA)~\cite{Brient:2004yq}. For optimal use 
of PFA, calorimeters needs to be homogeneous and finely segmented. 

\section{SDHCAL concept}\label{sec:sdhcal}

The SDHCAL prototype meets the ILC requirements by the combination of 
various technological choices. The homogeneity is achieved by the use 
of large Glass Resisitive Plate Chambers (GRPC) as the active 
medium combined with a power-pulsed embedded electronics. 
The power-pulsing suppresses the need of integrating a cooling circuit 
inside the detector by reducing the power consumption and heating. 
The homogeneity is further achieved by having all services 
(gas inflows and outflows, high voltage, data readout, etc.) 
to be on only one side of the GRPC and outside the HCAL. 

The fine segmentation is achieved transversally by the readout 
electronics system. An embedded Printed Circuit Board (PCB) with a checked 
side made of 1~cm$^2$ copper pads, reads the signal created by 
the passage of charged particles in the GRPC detector. 
The other side of the PCB holds the HARDROC ASICs~\cite{deLaTaille:2011zz}, 
that collect signals from the copper pads, digitizes them and 
transmits them to the outside data acquisition system. 
The ASIC provides 2-bit readout. 

The typical size of an avalanche inside the GRPC is 
around 1~mm$^2$. At high energy, the shower core is very dense, 
up to more than 100 avalanches per cm$^2$, 
and simple binary readout will suffer from saturation effect. 
Semi-digital readout (2-bit) reduces this effect and 
improves the energy reconstruction. 

The choice of this semi-digital scheme rather than the binary one 
was motivated by simulation studies~\cite{Laktineh:2011zz} and 
the three corresponding thresholds are used to tag pads fired by few, 
little or many crossing particles inside the shower. This choice
has been validated by results~\cite{CAN037} that will be 
discussed later in this article. 

\section{The GRPC cassette}\label{sec:cassette}

\begin{figure}[tbp] 
\begin{center}
\includegraphics[width=.95\textwidth]{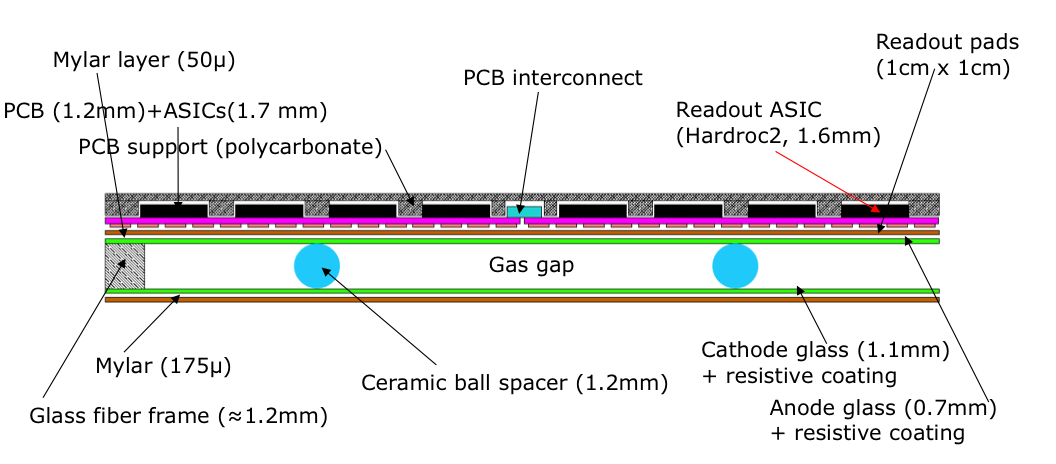}
\caption{Schematic cross-section view of a 1~m$^2$ GRPC cassette.} 
\label{fig:SDHCAL:Cassette} 
\end{center}
\end{figure}
 
A GRPC cassette~\cite{Laktineh:2011zz} has a 1~m$\times$1~m area 
and is 11~mm thick (see figure~\ref{fig:SDHCAL:Cassette}).
It contains one GRPC and its associated electronics. 
The cassette is a thin box consisting of two 2.5~mm thick 
stainless steel plates separated by 6~mm wide stainless steel 
spacers which form the walls of the box.
One of the two plates is 20~cm larger than the other to hold 
the PCBs used for the data acquisition as well as the gas outlets 
and the high voltage box. 
Precision machined stainless steel spacers, insulated from the GRPC, 
are making the cassette's sides. A polycarbonate mask is added around 
the ASICs to ensure that once the cassette is closed, the PCB is 
forced to stay into contact with the GRPC anode. This cassette 
structure ensures a homogeneous efficiency of the GRPC signal 
collection by the PCB copper pads.

The cassettes are inserted into a self-supporting 
structure~\cite{Laktineh:2011zz} with 50 13~mm thick spaces 
separated by 1.5~cm thick stainless steel plates.
The SDHCAL prototype is then a sampling calorimeter with 
2~cm thick absorber layers, nearly one radiation length, 
and 6~mm thick active detectors.

\subsection{GRPC}\label{sec:GRPC}
The GRPC is used in saturated avalanche mode~\cite{Ammosov:2007zz}:  
the avalanche is initiated by the crossing of the 1.2~mm thick 
gas gap by one or more charged particles. The gap is framed by two 
electrodes made of borosilicate float glass~\cite{Bedjidian:2010zz}. 
The anode and cathode thickness are 0.7~mm and 1.1~mm respectively. 
The smaller anode thickness enhances the signal in the copper 
pad closest to the crossing particle and lowers the relative signal 
seen by neighbouring pads. The high voltage used is typically 
7~kV. A glass fiber frame, width 3~mm, height 1.2~mm 
is used to seal the gas volume.

The gas distribution within the chamber is done through an 
L-shaped channel delimited by the chamber frame and a series of 
PMMA fibers~\cite{Bedjidian:2010zz}. 
Gaps between the fibers allow the gas to leave the channel at 
regular intervals, expanding into the main chamber volume. Under usual 
operating conditions the chamber volume is exchanged
every 20 minutes. This needs a gas flow of 3.6~l/h 
and an overpressure of 1~mbar in the chamber. This overpressure 
corresponds to a force per unit area of 100~N/m$^2$, which almost 
balances the attractive electric force between the plates.
The operating gas is a mixture of 93\% of TFE, 5\% of C0$_2$ and 
2\% of SF$_6$. TFE has been chosen for its low ionisation energy 
enabling efficient creation of avalanches. 
CO$_2$ and SF$_6$ are used as UV and electron quenchers respectively.
Studies of gas mixture performances are described in~\cite{Kieffer:2011zea}.

To maintain the gas gap over the entire chamber area, 
ceramic balls with 1.2~mm in diameter are glued to the cathode 
every 10~cm. A finite element analysis~\cite{Bedjidian:2010zz} taking 
into account the electric force and the glass plates weight has 
determined that for this distribution of balls, the maximum deflection 
of the anode glass is 44~$\mu$m. 
When the gas circulates, this deflection is reduced. To ensure the 
physical integrity of the chamber in case the gas circulates
but the high voltage is switched off, 13 ceramic balls are replaced 
for each m$^2$ by cylindrical glass disks glued on
both electrodes~\cite{Kieffer:2011zea}.
The glass electrodes are coated with a cheap bi-component painting
based on colloidal graphite. This painting can be silkscreen printed 
for a very good coating uniformity. In addition, the resistivity 
is adjustable by varying the proportion of the two components. The 
coating provides uniform spread of the high voltage over the 1~m$^2$ electrodes.

\subsection{Electronics}\label{sec:electronics}
A 1~m$^2$~GRPC is tiled with 6 PCB size of 
$\frac{1}{3} \times \frac{1}{2}$~m$^2$. The PCBs have eight layers.
On one external face, 1536 copper pads of $1\times 1$~cm$^2$ are 
printed. Copper pads are separated by 406~$\mu$m.
On the opposite face, 24 HARDROC2 ASICs~\cite{deLaTaille:2011zz} are 
soldered. Each ASIC is connected to $8\times 8$ pads through 
the PCB. The electronic channel cross-talk between two adjacent pads 
is less than 2\%~\cite{Bedjidian:2010yj}. 

Each of the 64 channels of an ASIC is made of a fast low-input-
impedance current-preamplifier. The gain of each preamplifier can be 
varied over 6 bits to correct for the non-uniformity between channels.
The 2-bit readout is done by 3 variable fast shapers (15-25 ns), each 
one followed by a low-offset discriminator.
Each fast shaper covers a different dynamic range leading to a total 
channel dynamic range from 10~fC to 15~pC. 

When the ASIC is in acquisition mode, the 192 discriminator outputs 
are checked each 200~ns. If one is fired, an event is stored in an 
integrated digital memory. This auto-triggering mechanism enables the 
operation of the calorimeter without external triggers. An ASIC event 
consists of the ASIC identification number, a 3-byte clock counter
and the output state of the 192 discriminators. The HARDROC2 can store 
up to 127 ASIC events. Once the memory is full, the ASIC
raises a signal and waits for instruction from the external Data 
Acquisition (DAQ) system. 

The ASICs are daisy chained on the PCB. Configuration data (the 64 
preamplifier gain, the 192 discriminator thresholds, etc.)
and the readout data are transfered through the chain of ASICs. The 
$\frac{1}{3} \times \frac{1}{2}$m$^2$ PCBs can be chained
themselves. On a 1~m$^2$~GRPC, the 6 PCBs are paired and 
connected to 3 DAQ interface boards (DIF) located on the larger 
cassette plate. This daisy-chaining reduces the amount of cables 
needed between the cassette and the external DAQ

An important feature of the ASIC is the possibility to be 
power-pulsed. It is possible to switch on/off the power-hungry 
parts of the ASICs by sending external control signal. 
The portion of the ASICs that should be power-pulsed is configurable 
before the acquisition starts. With everything on, the power 
consumption is 1.425~mW per copper pads. When most of the ASICs is 
switched off, it reduces to less than 0.2~muW. The ILC operating 
cycle~\cite{Adolphsen:2013kya}, will have a 200~ms period in 
which collisions occur only during 0.95~ms. By switching on the 
HARDROC2 only 0.5\% of the time, the SDHCAL power consumption is below 
10~$\mu$W per channel and  sufficiently low to avoid the need of an 
internal cooling system.

\section{Prototype assembly}\label{sec:prototest}
\subsection{The DAQ}\label{sec:daq}
DAQ is driven by the XDAQ software~\cite{Brigljevic:2003kg}.
The ASICs configuration data are stored in 
an ORACLE database and sent to the DIFs via HDMI cables.
Clock synchronisation is sent through the same HDMI cables.
The readout data are retrieved from the DIFs through USB cables. 
The use of USB concentrators and specific boards (DCC) to dispatch the 
HDMI signals allows to scale the DAQ from 1 cassette to the full 
prototype which comprises more than 460000 readout channels.

\subsection{Single cassette uniformity}\label{sec:effmult}
\begin{figure}[tbp] 
\centering
\includegraphics[width=.3\textwidth]{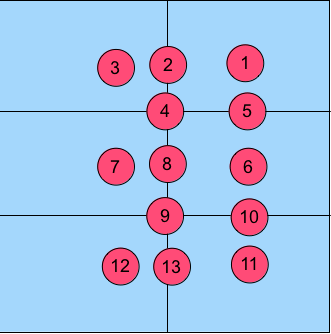}
\includegraphics[width=.3\textwidth]{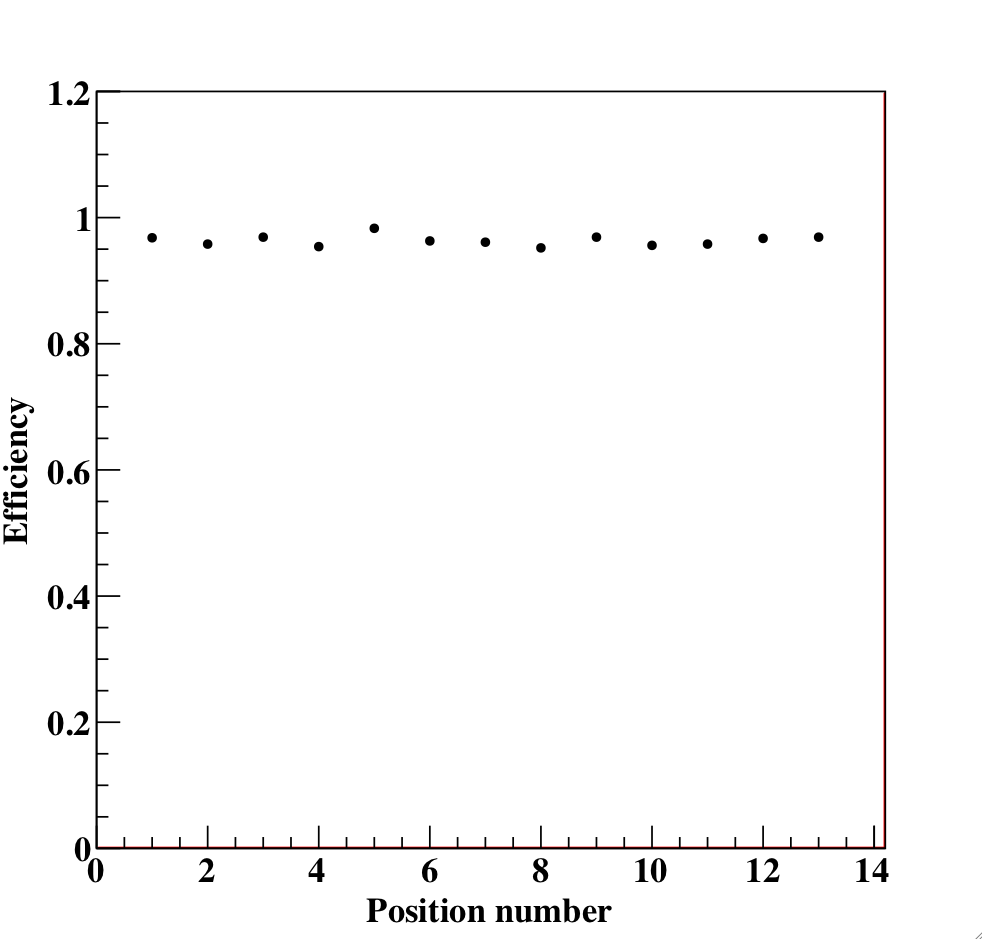}
\includegraphics[width=.3\textwidth]{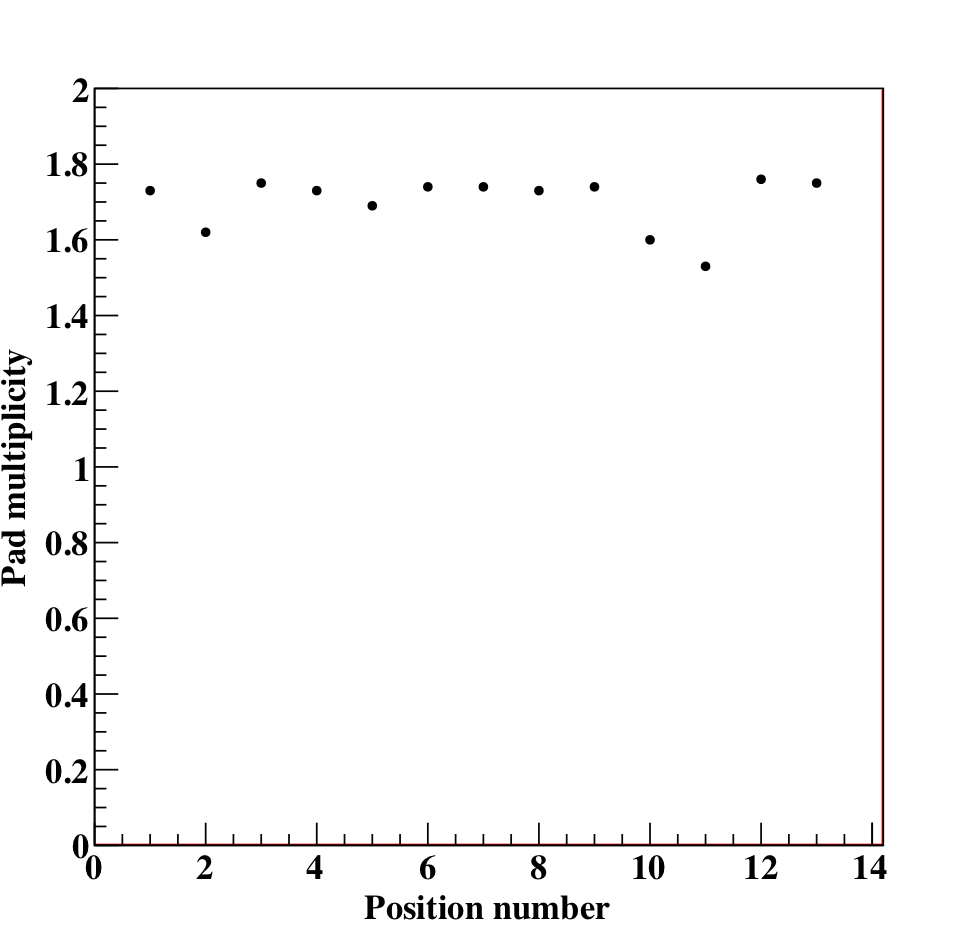}
\caption{Position (left) of the 13 locations where particle beams
were crossing a 1~m$^2$ chamber. Measured chamber efficiency (middle) 
and multipicity (right) for the 13 locations.}
\label{fig:chamberuniformity}
\end{figure}

A single GRPC cassette was tested in a test beam. The beam was sent
at various locations of the 1~m$^2$ cassette: at the center of different 
$\frac{1}{3} \times \frac{1}{2}$m$^2$ PCBs and at the junction of 
2 or 4 PCBs. The 13 locations where the beam was shot are shown in 
figure~\ref{fig:chamberuniformity}. The response of the chamber 
was checked for each of the 13 locations. The response consists of 
the efficiency and the multiplicity. The efficiency is the fraction
of crossing charged particles that fires at least one channel
at a distance of less than 3~cm around the expected position 
of the crossing charged particle. 
The multiplicity is defined as the mean number of fired channels 
when a crossing particle fires the chamber. 
Figure~\ref{fig:chamberuniformity} shows a very uniform response of the  
tested chamber.

\subsection{Prototype construction}\label{sec:construction}
For the construction, 10500 ASICs were tested and calibrated using a
dedicated robot that was used by CMS (yield : 93\% ). 
310 PCBs were produced, cabled and tested. 
They were assembled into sets of six to make 1~m$^2$ ASUs.
170 DIFs and 20 DCCs were built and tested.
50 detectors were built and assembled with their electronics embedded into 
cassettes. Cassettes were tested by sets of 6 using a cosmic test bench. The mechanical structure was built separately.
Full assembly of the structure, the cassettes and the HV and cooling 
services took place at CERN. 

\subsection{Prototype operation}\label{sec:operation}
The full prototype has then been exposed 
to pion and electron beams at the CERN SPS for 3 periods totalling 5 
weeks. For these beam tests, 48 GRPCs were inserted inside of the 
absorber structure. The 3 thresholds for the 2-bit readout were set to 
114 fC, 5 pC and 15 pC respectively, the average MIP induced charge 
being around 1.2 pC. 
The same electronics gain was used for all the channels and the 
high voltage applied on the GRPC was of 6.9 kV. The proportion of dead 
channels was about 1 per mil.

\section{Data analysis}\label{sec:analysis}
\subsection{Time analysis}\label{sec:time}
In the trigger-less data acquisition mode used for the test beam, 
all the activity in the detector is recorded: 
isolated hits (fired pads) due to noise, particles crossing the detector or 
showering in it. The electronics runs with a clock tick of 200~ns.
For the vast majority of clock ticks, there are no hits 
recorded in the prototype. Analysis of the random hit distribution
leads to a mean noise estimation of 0.35 hits per 200~ns for 
the more than 460000 readout channels~\cite{CAN037}. 
Potential physics events are reconstructed by combining hits 
from 3 consecutive clock ticks if the middle clock tick has at 
least 7 hits.

\subsection{Data quality}
To monitor the calorimeter performance, the efficiency and 
multiplicity of the GRPC response are estimated using beam of muons.
For each layer, muon tracks are reconstructed from the hits of the 
other layers. The expected impact point of the track in the layer
under study is then used to compute efficiency and multiplicity for the
chamber as described in section \ref{sec:effmult}. 
The procedure can be done to monitor a full chamber or any portion 
of it down to pad by pad monitoring. For the test beams, the 
number of recorded beam muon events was sufficient to monitor 
the calorimeter at the ASIC level (a group of 8 by 8 pads).
The efficiency has been measured to be 95\% with a 6\% dispersion 
between ASICs. The multiplicity has been measured to be 1.73 hits
per crossing particles with a dispersion of 0.25 hits between ASICs.

\subsection{Efficiency modeling}
\begin{figure}[tbp] 
\centering
\includegraphics[width=.45\textwidth]{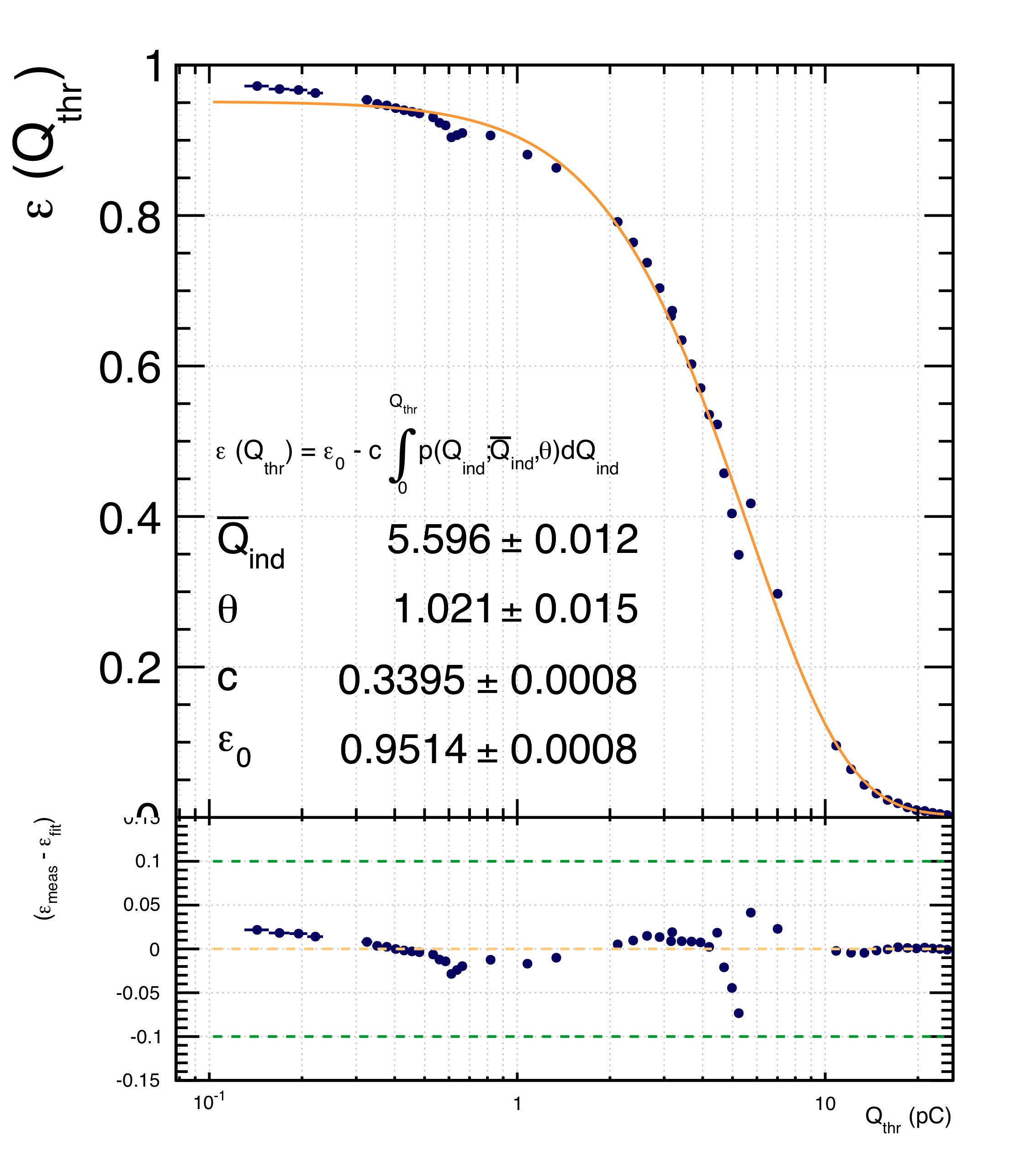}
\includegraphics[width=.45\textwidth]{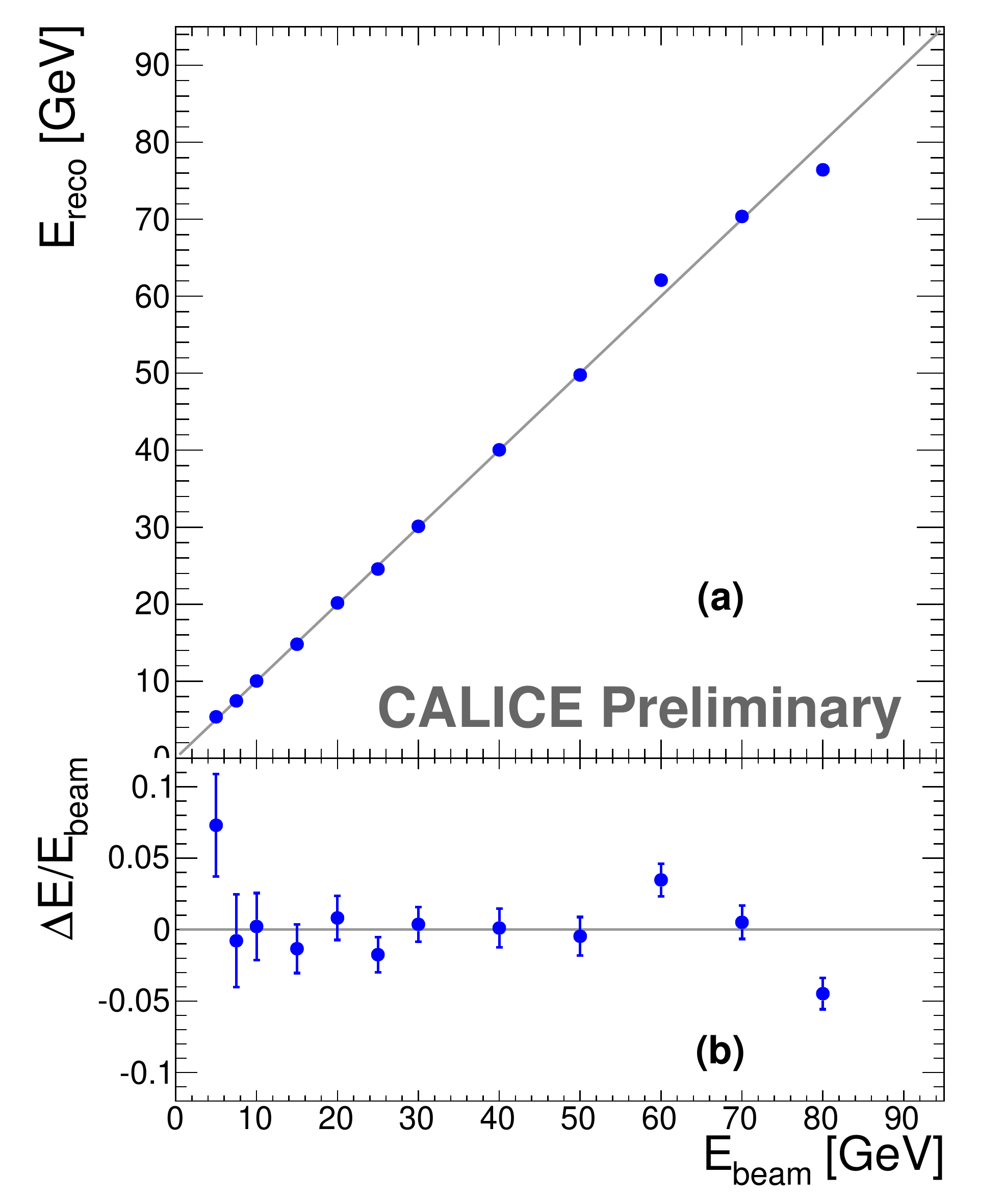}
\caption{Top left: efficiency versus threshold for data 
(blue points) and model (orange line). 
Bottom left: efficiency difference between data and model as a 
function of the charge threshold.
Top right: mean reconstructed energy for pion showers as a function 
of the beam energy. 
Bottom right: relative deviation of the pion mean reconstructed energy 
relative to the beam energy as a function of the beam energy.}
\label{fig:results}
\end{figure}

During special runs, the thresholds were varied in a few layers 
to measure the efficiency as a function of the threshold as 
shown in figure~\ref{fig:results} left. The  charge induced by an 
avalanche can be modelled by a Polya distribution 
(eq.~\ref{eq:polya}). The efficiency is modelled as the combination 
of the probability $\epsilon_0$ for a crossing
charged particle to initiate an avalanche and the probability that 
the induced charge is above the threshold. This leads to the modelling
of equation~\ref{eq:effvsthreshold} with the values of 
the model parameters shown on figure~\ref{fig:results} left.
\begin{eqnarray}
\label{eq:effvsthreshold}
\epsilon (Q_{thr}) &=&  \epsilon_0 - c \int_0^{Q_{thr}}
p(q;\theta,\bar{Q}_{ind}) dq \,,
\\
\label{eq:polya}
\mathrm{with} \, p(q;\theta,\bar{Q}_{ind}) &=& 
\left( q \frac{1+\theta}{\bar{Q}_{ind}} \right)^\theta 
\exp \left\lbrace - q \frac{1+\theta}{\bar{Q}_{ind}} \right\rbrace \,,
\end{eqnarray}

\subsection{Pion energy reconstruction}
A simple selection of showers due to interacting pions is done for energy reconstruction: 
electrons are rejected by requiring that the shower starts after the
fifth layer or that more than 30 layers have at least 4 hits. Muons
are rejected by asking that the mean number of hits per fired layers 
should be above 2.2 and that at least 20\% of the fired layers
have a spatial hit distribution with a RMS above 5~cm. Neutral 
particles are rejected by requiring the presence of at least 4 hits
in the first 5 layers. Details of the selection are described 
in~\cite{CAN037a}

The energy of a pion is reconstructed as 
\begin{equation}
\label{eq:Ereco}
E_{reco}=\alpha(N_{tot}) N_1 + \beta(N_{tot}) N_2 + \gamma (N_{tot}) N_3
\end{equation}
where $N_{tot}=N_1+N_2+N_3$ and $N_1$, $N_2$ and $N_3$ are the 
exclusive number of hits associated to the first, second and third 
threshold, and, $\alpha$, $\beta$ and $\gamma$ are quadratic 
functions. The reconstructed energy as a function of the beam energy 
is shown on figure~\ref{fig:results} right. The calorimeter response
is linear with deviations from linearity below 5\%
for pion energies between 7~GeV and 80~GeV. The corresponding resolution 
decreases from 25\% for 5~GeV pions down to 8.9\% for 80~GeV 
pions~\cite{CAN037a}. Further improvements on the energy resolution
are shown in~\cite{CAN037b}.

\section{Outlook} 
A prototype of a semi-digital imaging calorimeter has been build
using GRPC as sensitive detectors. The GRPC detectors have proven to 
be cheap, homogeneous, thin, 
with very low noise and very few dead zones. 
The GRPCs are associated with an embedded readout electronics and 
allow a very fine segmentation of the readout of 9216 channels per 
m$^2$. 
Further developments are ongoing on the software reconstruction. 
For example, the use of Hough Transform to improve tracking and 
energy reconstruction~\cite{Procedding:TIPP2014}. Hardware 
developments are ongoing as well to increase the size of the 
chamber to 2~m$^2$ which implies to revisit the gas distribution 
inside the chamber and the ASICs daisy chaining.

\acknowledgments

This work has been done within the SDHCAL group of the CALICE 
collaboration. The SDCHAL group comprises teams from 
IPNL (Lyon, France), LLR (Palaiseau, France), LAPP (Annecy, France),
LPC (Clermont-Ferrand, France), OMEGA (Orsay, France), 
CIEMAT (Madrid, Spain), UCL (Louvain, Belgium),
Universiteit Gent (Belgium), Tsinghua University (Beijing, China), 
Universit\'e de Tunis (Tunisia).

\end{document}